\documentclass[conference]{IEEEtran}
\IEEEoverridecommandlockouts
\usepackage[
  style = ieee
]{biblatex}
\usepackage{amsmath,amssymb,amsfonts}
\usepackage{algorithm}
\usepackage{algpseudocode}
\usepackage{graphicx}
\usepackage{textcomp}
\usepackage{xcolor}
\def\BibTeX{{\rm B\kern-.05em{\sc i\kern-.025em b}\kern-.08em
    T\kern-.1667em\lower.7ex\hbox{E}\kern-.125emX}}
\begin{document}

\title{Auctioning Escape Permits for Multiple Correlated Pollutants Using CMRA
}







\author{\IEEEauthorblockN{Keshav Goyal\IEEEauthorrefmark{1},
Sooraj Sathish\IEEEauthorrefmark{2} and Shrisha Rao\IEEEauthorrefmark{3}}
\IEEEauthorblockA{IIIT Bangalore\\
Email: \IEEEauthorrefmark{1}keshav.goyal@iiitb.ac.in,
\IEEEauthorrefmark{2}sooraj.sathish@iiitb.ac.in,
\IEEEauthorrefmark{3}srao@iiitb.ac.in}}

\maketitle

\begin{abstract}
In the context of increasingly complex environmental challenges, effective pollution control mechanisms are crucial. By extending the state of the art auction mechanisms, we aim to develop an efficient approach for allocating pollution abatement resources in a multi-pollutant setting with pollutants affecting each other's reduction costs. We modify the Combinatorial Multi-Round Ascending Auction \cite{cmra}\cite{dotecon} for the auction of escape permits \cite{sui1}\cite{sui2} of pollutants with co-dependent reduction processes, specifically, greenhouse gas emissions and nutrient runoff in Finnish agriculture. We show the significant advantages of this mechanism in pollution control through experiments on the bid prices and amount of escape permits sold in multiple auction simulations.
\end{abstract}


\section{Introduction}
As the world becomes more industrialized and our population continues to grow, we are witnessing an increase in the complexity of environmental issues ranging from air and water pollution to climate change. Pollution control is a critical aspect of environmental conservation through strategies that aim to reduce the emission of pollutants into the environment. An effective and economic strategy is the use of auctions for pollution permits, also known as escape permits \cite{sui1}\cite{sui2}. These permits allow individuals and organizations to emit a certain amount of pollution, thus limiting the total pollution and driving research in emission reduction. 

In a multi-pollutant setting, the escape permit auctions aim to allocate permits for different pollutants in a single auction. This is done to incorporate the economic and environmental gains or losses from multi-pollutant reduction technologies \cite{Gao2021} which might decrease or increase the cost of reduction for multiple pollutants. Thus, by avoiding the assumption that any 2 pollutants are uncorrelated in their reduction costs, we are able to model the real world context. However, the multi-pollutant setting carries an added level of complexity as compared to independent pollution types. For example, for a particular agent, reducing the emissions of one pollutant might lead to a nonlinear reduction in the emissions of another pollutant thus reducing the total cost of reduction for both the pollutants.

Bagchi et al. \cite{bagchi} present a solution for auctioning escape permits in a single pollutant setting. The study proposes two solutions, a reverse auction protocol and a forward auction protocol, to solve the case of an enterprise allocating pollutant reductions among its autonomous subsidiaries. The solutions proposed in their work are based on the Vickrey-Clarke-Groves (VCG) auction mechanism. Another work in the single-pollutant domain is that of Khezr and MacKenzie \cite{Khezr2021-wd} which proposes a modification to the uniform-price auction for the efficient allocation of pollution permits through truthful bidding. However, in the multi-pollutant setting where emission enterprises will have various combinations of pollutants to discharge in the production process, Guo et al. \cite{smra} show that single-pollutant auction mechanisms are no longer applicable.

Guo et al. \cite{smra} propose a modification of Simultaneous Multi-Round Ascending auction (SMRA) for auctioning escape permits in a multi-pollutant setting. This study by Guo et al. represents the only work that specifically addresses the multi-pollutant setting. SMRA is a multi-item combinatorial auction where the buyers can choose their own combination of emission rights. This research focuses on the combined transactions for emission rights of international carbon sequestration and other pollutants in forestry. The feasibility and applications of SMRA auction are explored through 4 experiments conducted by human subjects acting as buyers and sellers in the simulated auction. Although SMRA is popularly used worldwide, Bedard et al. \cite{smrabad} show that even a classical First-Price Sealed-Bid (FPSB) auction performs better than SMRA in terms of efficiency, revenue generated and winner's curse. SMRA also does not achieve effective allocation when the escape permits are complementary (reducing one pollutant increases another). Another popular multi-item auction mechanism is the Combinatorial Clock Auction (CCA) \cite{ccadef}. However, CCA faces various objections from the bidders. Janssen and Kasberger \cite{Cca} show that under certain conditions, CCA does not achieve equilibrium and the bidders are left dissatisfied. These shortcomings render CCA incompatible for escape permits auctions as it is essential to have willingness to participate and low price bids to incentivize pollution reduction.

Our solution is to use the Combinatorial Multi-Round Ascending Auction \cite{dotecon}\cite{cmra} to address the allocation of multiple different escape permits to autonomous agents. The CMRA auction was pioneered by the Danish Energy Agency in 2016 \cite{dotecon} and is based on CCA while also aimed at covering its shortcomings. Applications of CMRA are limited to spectrum auctions where spectrums with similar bandwidths affect the utility of the buyer. In spectrum auctions, CMRA solves the problem of direct correlations between the utilities of different spectrum groups. A buyer would be willing to pay a higher cost for grouped spectrums having similar bandwidths than ones with unrelated bandwidths. Thus, the process for determining utilities is straightforward. This is the first work to extend CMRA in the domain of pollution reduction. We extend the CMRA mechanism to the case where utilities of two escape permits are indirectly related by their reduction processes and costs. Our modified CMRA mechanism applies to the auction of escape permits where the utility for the escape permits is determined indirectly from the cost of reduction of the pollutants. The auction involves multiple rounds with each agent able to make multiple bids in each round. In each round, the agents independently decide the number of escape permits for each pollutant involved in a bid and the price of the bid according to their cost functions. At the end of the auction, every agent is allocated a certain combination of escape permits for the different pollutants. An allocation mechanism ensures that no agent ends up with an allocation that they do not bid for. Kasberger and Teytelboym \cite{cmra} study the properties of CMRA and show that it can achieve equilibrium in cases where CCA truthful bidding might fail. Kasberger and Teytelboym \cite{cmra} also highlight that CMRA might lead to reduced revenue under certain conditions, however, this does not pose a problem for our use case as for the multi-pollutant reduction setting, reaching an equilibrium is more important than generating maximum revenue.
 
 We consider the setting of multiple pollutants where the reduction of a pollutant can affect the reduction of other pollutants linearly. An efficient and economical mechanism is proposed for the problem of allocating multi-pollutant escape permits having correlated costs to different agents. We perform multiple simulations modeling the reduction of greenhouse gas emissions and nutrient runoff emissions due to agricultural activities. Lötjönen and Ollikainen \cite{2019multiplepollutant} construct cost functions for reducing greenhouse gas emissions and nutrient runoff in Finnish agriculture for crop and dairy farming. Through the use of real world data, they focus on determining the cost of reduction of GHG pollution while taking into account the benefits of reducing nutrient runoff pollution and vice versa. The GHG and nutrient runoff reduction costs for our simulations are derived from the results obtained by  Lötjönen and Ollikainen \cite{2019multiplepollutant}. We perform experiments to determine the Transaction Proportion and Transaction Price achieved by implementing our mechanism to the above setting. We show the high willingness to participate and the incentive to reduce pollutants achieved by our mechanism.

 Our contributions are:
 \begin{itemize}
     \item In Section \ref{AA} we present an extension of the CMRA auction mechanism for the correlated multi-pollutant setting. The mechanism uses the pollutant reduction costs as the cost functions for the escape permits. It works with a combined cost curve of all the pollutants instead of individual cost functions for each pollutant. To the best of our knowledge, this is the first work that discusses the application of CMRA mechanism in the auction of escape permits for pollution reduction.
     \item In Section \ref{sec:exp1} we show the flexibility of our mechanism as it does not necessitate the sale of all the escape permits under auction. This is a highly desirable property as it incentivizes autonomous agents to reduce their emissions instead of buying a lot of escape permits. A significant advantage of this flexibility is the opportunity to implement the ``safety valve" mechanism \cite{saftey} without exceeding the permit cap, serving as a safeguard against unforeseen spikes in the cost of pollution reduction. Thus, we can achieve pollution reduction both economically and environmentally.
     \item In Section \ref{sec:exp2} we show the low costs incurred by the agents who have a higher correlation between the reduction of two pollutants. This is an essential property of the auction mechanism as it incentivizes the research and use of better pollution reduction technologies. Reducing costs for the agents enhances their willingness to participate, fostering an economically efficient allocation of permits and achieving reduction goals.
 \end{itemize}

\section{CMRA auction for pollution reduction permits}
\subsection{CMRA Auction Architecture}
\label{algos}
We first discuss the CMRA auction mechanism and its salient features. The following terminology is followed:
\begin{itemize}
    \item {\bf Goods:} There are \(m\) goods \(\{G_i \vert \hspace
{0.2cm} 1 \leq i \leq m\}\) to be auctioned. The total amounts to be auctioned are represented as a single vector of length \(m\) : \(\{g_1, g_2,..., g_m\}\).
    \item {\bf Bidders:} There are \(n\) autonomous bidders \(\{B_i \vert \hspace
{0.2cm} 1 \leq i \leq n\}\) with their private utility functions \(U_i(x\vert G_j)\) which map the maximum value that they can pay for an amount \(x\) of the good \(G_j\). \(U_i(x)\) is the vector of utilities for the vector \(x\) representing amounts of all the goods.
\end{itemize}

The auction consists of multiple rounds of bidding. Each round has an associated price vector (clock price \(p\)) of size \(m\) which contains the prices of all goods. In each round a bidder can make two types of bids:
\begin{itemize}
    \item \textbf{Headline Demand: }At every clock price \(p\), a bidder \(B_i\) submits exactly one headline demand bid vector \(h_i(p) \in [0,1]^m\). This represents how much of each good the bidder wants to buy at the current clock price. If the total demand for any good is more than the supply, its clock price increases in the next round. Thus, If for any \(G_j\), \[\sum_i^n {[h_i(p)[j]]} > 1\] \(p[j]\) is increased for the next round.

    \item \textbf{Additional Bids: }This is a novelty of CMRA. A bidder can make bids for different amounts at their chosen prices (under a few constraints). The additional bid by a bidder \(B_i\) for an amount vector \(x\) will be a vector \(A_i(x)\) of size \(m\) representing the price at which \(B_i\) wants to buy \(x\). Some constraints are:
    \begin{itemize}
        \item If \(x = [0]^m\), then \(A_i(x) = [0]^m\)
        \item \(\forall j \leq m\), \(A_i(x)[j] < x[j].c[j]\) 
        \item Additional constraints of CMRA ensure that additional bids do not deviate too much from the headline demand.
    \end{itemize}
\item \textbf{Closing Rule: }The allocation is done by picking exactly one bid from each bidder such that maximum revenue is generated. This could be a headline demand or an additional bid. If such an allocation is not possible, the auction continues. Moreover, the revenue generated by excluding any bidder must not exceed the maximum allocation's revenue.
\end{itemize}

Algorithm \ref{algo} is used for calculating headline demand and additional bids. In lines 2 and 3, each bidder, with a utility function $(U_i)$, calculates their headline demand $(h_i(p))$ at a given clock price $(p)$ using the inverse of their utility function. In line 5, additional bids $(A_i(x; p))$ are calculated for quantities less than the headline demand. The bid function $(B_i(; p))$ maps different amounts of the goods to the final bids submitted by the agent for the current round (clock price \(p\)). This may also contain bids from previous rounds. In line 6, the bid functions $(B_i(; p))$ are updated to reflect the maximum value between previous bid functions and these newly calculated additional bids.

\begin{algorithm}
\label{algo}
\caption{Calculate Headline Demand and Additional Bids}
\begin{algorithmic}[1]
\Require{Bidders: List of bidders with Utility functions \(U_i\)}
\Require{Clock Price: \(p\)}
\Ensure{Bid functions at clock price \(p\) for all bidders: $B_i(·; p)$}
\Procedure{BidFunctions}{$p, U_i$}
\For{each $i$} \algorithmiccomment Iterate over all Bidders
    \State $h_i(p) = U_i^{-1}(p)$ \algorithmiccomment Calculate headline demand
    \For{each $ x < h_i(p)$}
        \State $A_i(x; p) = U_i^{-1}(x)$ \algorithmiccomment Find Additional bids
        \State $B_i(x; p) = \max(B_i(x; p), A_i(x; p))$ \algorithmiccomment Update
        
    \EndFor
\EndFor
\State \textbf{Return} Bid functions: $B_i(·; p)$ for each bidder
\EndProcedure
\end{algorithmic}
\end{algorithm}


\begin{algorithm}
\label{algo2}
\caption{Auction}
\begin{algorithmic}[1]
\Require{Bidders: List of bidders with Utility functions \(U_i\)}
\Require{Total Goods: Vector of total quantities of goods \(g\)}
\Require{Initial clock price: $p_{\text{initial}}$}
\Ensure{Bid functions at clock price \(p\) for all bidders: $B_i(·; p)$}
\Procedure{Auction}{$\text{Bid functions}, p, \text{Bidders}$}
\State $MaxBid = -Infinity$ \algorithmiccomment Initialize\\
\text{    \# Find maximum allocation}
\For{ all allocations $(x_1, x_2, .., x_n) \in [0^{\vert g \vert}, g]^n$}
\If{ $MaxBid <  \sum_{i = 1}^n(B_i(\tilde{x}_i; p))$ }
    \State $MaxBid =  \sum_{i = 1}^n(B_i(\tilde{x}_i; p))$
    \State $x = (x_1, x_2, ., x_n)$
\EndIf
\EndFor
\If{MaxBid is feasible}
    \State \textbf{Return} $x$ \algorithmiccomment End auction with allocation $x$
\Else \\
    \text{        \# Increase the clock price}
    \State $p = p + \delta$ \algorithmiccomment$\delta$ is a small increment
\EndIf
\EndProcedure
\end{algorithmic}
\end{algorithm}

Algorithm \ref{algo2} outlines the process of conducting an auction. It starts with a list of bidders with cost functions, total goods, and an initial clock price. In line 2, The bid functions are first initialized for all the agents. In each round the maximum bid $(MaxBid)$ for all possible allocations is calculated as shown in lines 3-9. In line 10, If the $MaxBid$ is feasible, the auction ends with the allocation. Otherwise, the clock price is increased by a small increment in line 13, and the process repeats. It adjusts the clock price based on the bids and continues until a feasible allocation is found. This ensures that the auction process is dynamic and responsive to the bids made by the participants.

\subsection{Extended CMRA for Escape Permits}
\label{AA}
We now show how the CMRA auction mechanism is extended to our problem of allocating escape permits for multiple codependent pollutants. As the problem is modeled as an auction of escape permits, an agent whose cost of reduction for \(x\) amount of pollutant is high, would like to bid more for an escape permit of \(x\) amount for the pollutant. Conversely, if the agent can reduce emissions at a lower cost, it would prefer to do that over bidding high on escape permits. The primary aim of the auction is to create an efficient equilibrium over the escape permits and also introduce an escape permit cap. 
The inputs to the extended auction mechanism are:
\begin{itemize}
    \item \textbf{Pollutants: }There are \(m\) pollutants \(\{P_i \vert \hspace
{0.2cm} 1 \leq i \leq m\}\). Any two of these pollutants might be co-dependent or independent concerning emission patterns and reduction costs.
    \item \textbf{Bidders: }There are \(n\) bidders, \(\{B_i \vert \hspace
{0.2cm} 1 \leq i \leq n\}\). Each bidder has its private cost function \(C_i\). So, for bidder \(B_i\), cost of reducing pollutant \(P_j\) independently by \(x\) amount is \(C_i(x\vert p_j)\). Whereas the combined cost function for all pollutants is \(C_i(x)\) where \(x\) is the vector of amounts of all pollutants. The combined cost function is a result of the codependence between the reduction costs of different pollutants. Thus, the relation 
    \[C_i(x) = \sum_j C_i(x\vert P_j)\]
    is not guaranteed.
    \item {\bf Escape} {\bf Permits:} There are \(m\) escape permits \(\{E_i \vert \hspace
{0.2cm} 1 \leq i \leq m\}\). Each escape permit is a single unit indivisible good in the auction. The relation between two escape permits depends on the relation between corresponding pollutants. For example, if two pollutants are substitutes of each other (reducing one also reduces the other pollutant), then their escape permits are also substitutes of each other. The total amounts to be auctioned (escape permit caps) are represented as a single vector of length \(m\) : \(\{e_1, e_2,..., e_m\}\).
\end{itemize}
    Then, for extending the CMRA auction, we define the utility of every bidder \(B_i\) for 1 unit of a permit \(E_j\) as the cost of reduction of 1 unit of pollutant \(P_j\) for that bidder.
    \[U_i(x \vert E_j) = C_i(x \vert P_j)\]
    and
    \[U_i(x) = C_i(x)\]

Some of the features of this CMRA process in the context of escape permits are:
\begin{itemize}
    \item A bidder will not get any escape permit only if they do bid for zero escape permits.
    \item A bidder will never get a combination of permits that they did not bid for.
    \item The codependence of escape permits is taken care of as the bidders themselves create their combinations.
    \item The bidders do not have the incentive to falsify their headline demands as they can always make additional bids at lower prices in later rounds.
    \item Extended CMRA ensures either efficient allocation of permits (only the parties with a high cost of reduction get more permits) or it ensures an allocation that guarantees cheap escape permits to everyone but fewer permits issued.
\end{itemize}
\section{Experiments and Results}\label{section:experiments}
We want to study the advantages of using CMRA auction in a multi-pollutant setting involving nutrient runoff and GHG emissions in Finnish agriculture for both dairy and crop farming. We quantify these benefits by noting the transaction price and the transaction proportion of multiple auction instances of the extended CMRA mechanism in the setting.

For conducting the experiments, the extended CMRA mechanism from Section \ref{AA} is implemented using the algorithms discussed in Section \ref{algos}. Since we are working in the truthful bidding paradigm of the auction mechanism,
\begin{itemize}
    \item For each agent, we need to calculate all possible bids having a non-negative surplus.
    \item In the Closing Rule, we need to consider all possible combinations of different bids to find the maximum allocation.
\end{itemize}
Thus, the computational time complexity of one round of the extended CMRA mechanism is calculated to be $\mathcal{O}(e_{\tau}^{m.n})$ where \(n\) is the number of agents, \(m\) is the number of different escape permits (or number of pollutants) and \(e_\tau\) is the maximum value among the different escape permit caps.
\[\tau = \underset{i}{\arg\max} (e_i)\]
This result is supported by the work by Lehmann et al. \cite{10.7551/mitpress/9780262033428.003.0013} showing that the winner determination problem in combinatorial auctions is $\mathcal{NP}$-hard. Due to the high complexity, for experimental purposes, the implemented simulation assumes 2 agents and 2 codependent pollutants (GHG emissions and nutrient runoff). The 2 agents can be any combination of crop farmers and dairy farmers (having different or slightly similar reduction cost functions).

As shown in section \ref{AA}, the utility of an agent for \(x\) units of an escape permit in the extended mechanism is the same as the cost of reduction of \(x\) units of corresponding pollutant for the agent. The reduction costs are represented by cost functions \((C_i(x\vert P_j))\) which map an amount \(x\) of the pollutant \(P_j\) to its cost of reduction for bidder \(B_i\). The cost functions used for our experiments are based on the results obtained by Lötjönen and Ollikainen \cite{2019multiplepollutant}. Their work resulted in the development of aggregated cost functions for the reduction of GHG and nutrient runoff in Finnish agriculture while accounting for their codependence. As the obtained functions are aggregated over multiple data points, instead of using the cost functions at face value, we add Gaussian random noise to the functions and interpolate them to be assigned to the agents in different instances of the simulation.

Using the implemented mechanism under the above assumptions, we calculate the transaction price and transaction proportion over 50 independent auction instances. The insights gained in Sections \ref{sec:exp1} and \ref{sec:exp2} conclude the benefits of applying the extended CMRA mechanism to the problem of multi-pollutant escape permit auction.
\subsection{Transaction Proportion}
\label{sec:exp1}
\begin{figure}[h]

\centering
\includegraphics[width=0.45\textwidth]{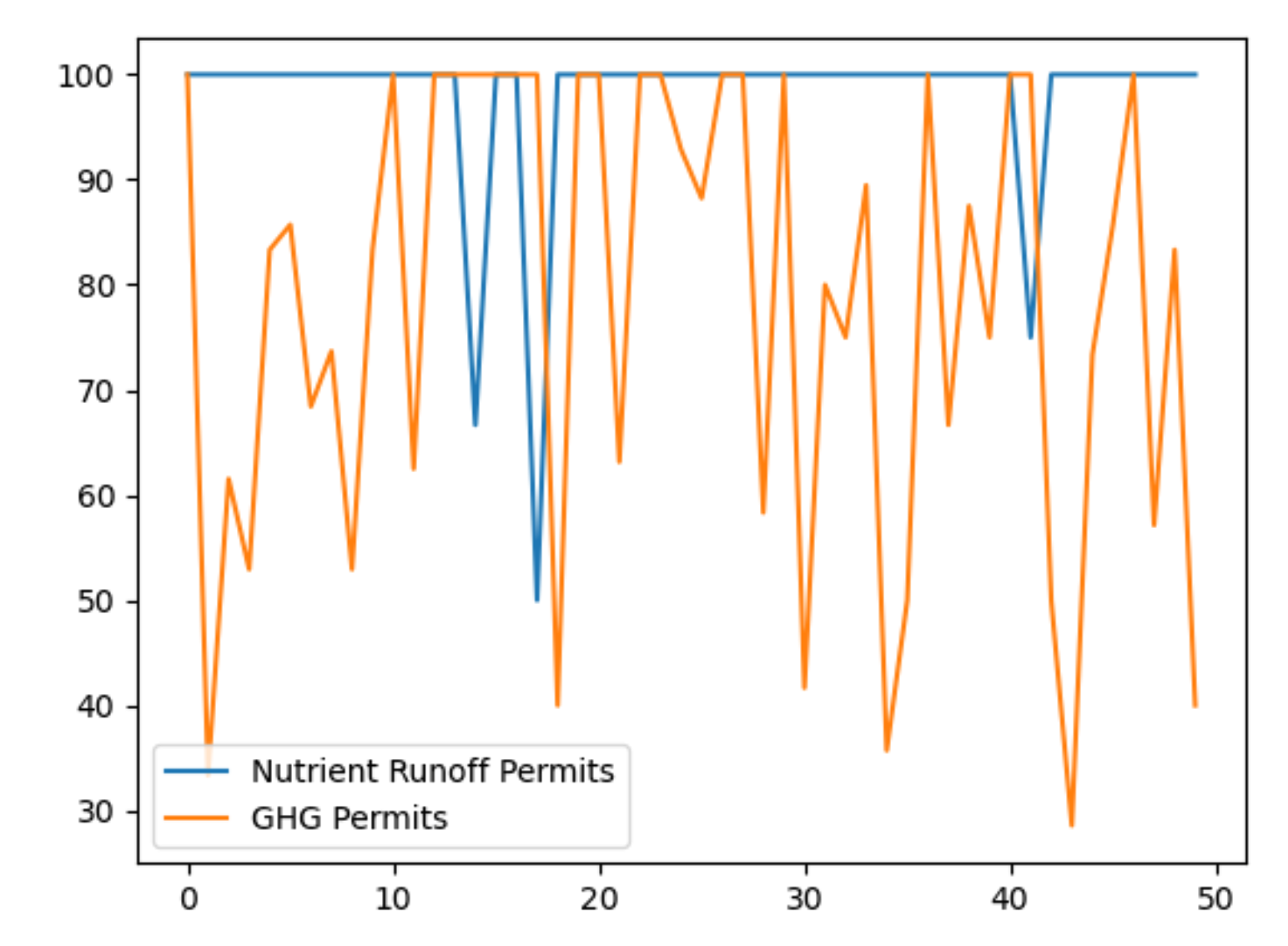}
\caption{Transaction Proportion Analysis}
\label{tprop}
\end{figure}

Transaction proportion is the percentage units of each pollutant sold from the total available with the auctioneer. Figure \ref{tprop} is the comparison of transaction volume, reflecting the proportion of the number of each auction item in all transactions to the total amount to be traded. A high transaction proportion indicates the success of an auction and also reflects the willingness to participate in the agents. A higher transaction proportion means that a higher quantity of escape permits are sold to the agents and thus they will be willing to participate.

We note the observation that in almost all auctions more than 30\% percent of GHG escape Permits are unsold. The reason for this is that on average, the cost of reduction of 1 unit (kgNe) of nutrient runoff is higher than that of 1 unit (kgCO2e) of GHG. So the agents would prefer to reduce the GHG emissions (consequently reducing nutrient runoff) instead of buying escape permits for GHG. For nutrient runoff, the agents would rather buy the escape permits than try to reduce them through higher costs.

 Classical auctions like VCG or English auctions would almost always lead to 100\% transaction proportion, all the permits need to be sold. Whereas, in the extended CMRA auction, the above observation highlights its flexibility wherein the agents are encouraged to reduce the pollutant instead of buying its escape permits if the auction price exceeds their cost of reduction. The unsold permits will allow the seller enterprise to implement the ``safety valve" mechanism \cite{saftey} without exceeding the escape permit caps. Under a ``safety valve" mechanism, the enterprise, seeking to reduce emissions, commits to selling escape permits to its autonomous subsidiaries upon request at a predetermined fixed price. Usually, safety valve permits are issued over and above the escape permit caps which poses an environmental problem. This is considered a major shortcoming of the safety valve mechanism \cite{edfWhatsafety}\cite{Gilbert2008TRADABLEPP}. However using the unsold permits of the extended CMRA mechanism, there is room for safety valve permits in the current escape permit allowance.


\subsection{Transaction Price}
\label{sec:exp2}
Transaction price signifies the average spending of a buyer at the end of the auction. It is desirable to have low spending in the auction for the agents. For an agent, a lower cost for each unit of escape permits means a higher surplus. Precisely, the transaction price is measured as:
\[ \frac{BidPrice}{TotalBidUnits} \]

\begin{figure}[h]

\centering
\includegraphics[width=0.45\textwidth]{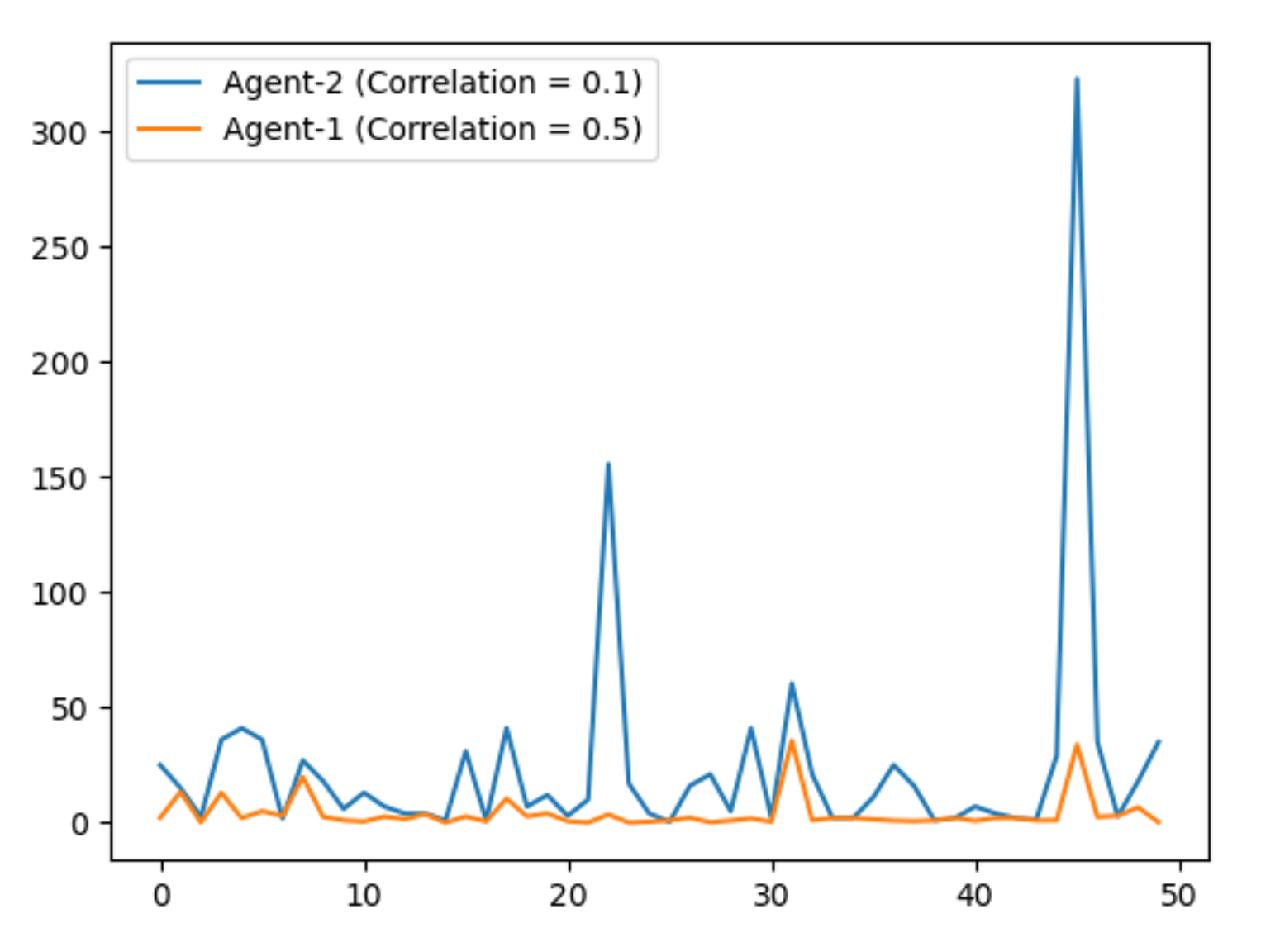}
\caption{Transaction Price for 2 agents with correlation coefficients 0.5 and 0.1 respectively}
\label{t_pric}
\end{figure}

We compare the transaction prices across auction instances for 2 agents with different correlation coefficients in Figure \ref{t_pric}. This means that the correlation between the reduction cost of the 2 pollutants is different for the 2 agents.

We note the observation that agent 2 with a correlation of 0.1, has a much higher avg. price as compared to agent 1 with a correlation of 0.5 . The reason for this is that on average, agent 2 has to buy reasonable amounts of both the permits including the costlier nutrient runoff permits. This is in effect due to the low-correlation coefficient of agent 2 as it can reduce only 0.1 units of nutrient runoff on reducing 1 unit of GHG emissions. Whereas agent 1 can reduce nutrient runoff at a lower cost by reducing GHG emissions using the high correlation coefficient of 0.5 . Thus agent 1 need not buy the escape permits for nutrient runoff as it will be reducing it through GHG reduction.

In a single-pollutant auction mechanism, the correlation coefficient doesn't factor in which would lead to a higher transaction price as compared to CMRA irrespective of the correlation between the reduction of pollutants. Whereas, the extended CMRA auction is able to benefit those agents who have a high correlation in the cost of reduction of the two pollutants. This is an overwhelming incentive to deploy multi-pollutant reduction technologies \cite{Gao2021} and invest in its research and development as they will inadvertently lead to higher correlation coefficients. Another result of the low transaction price is the increased willingness to participate in the agents as they can extract economic outcomes from the auction process. Thus, the modified CMRA mechanism helps both economically and environmentally when the reduction mechanisms of the pollutants are related.



\section{Conclusion}
In this work, we have extended the CMRA auction mechanism for the auction of multiple correlated pollutants using escape permits as auction goods. This is done by indirectly determining the utility functions of each agent using their pollution reduction costs. We have presented the implications of utilizing the CMRA auction for pollution reduction through autonomous agents. We utilize the truthful bidding paradigm of the auction and highlight the major benefits gained by the CMRA mechanism in the sector of crop and dairy farming for the reduction of GHG emissions and nutrient runoff pollution.

We implement the extended CMRA auction mechanism for a simulated setting in the Finnish agriculture sector and show how the extended CMRA auction mechanism would have a high willingness for individuals and enterprises to participate in the auction as opposed to classical auction mechanisms. Transaction proportion reveals that with unsold permits acting as a buffer, our extended mechanism addresses a major shortcoming of the safety valve mechanism \cite{saftey} and makes it economic in a practical context without exceeding the permit caps. We show through an analysis of transaction price that the mechanism gives crucial importance to the correlation between the reduction costs of pollutants, thus encouraging multi-pollutant reduction technologies. Thus the proposed CMRA algorithm modified for the multi-pollutant framework is beneficial to the environment as it drives pollution reduction and research in multi-pollution reduction technologies while being economic for the pollution emitters.
\printbibliography

\end{document}